\documentclass[conference]{IEEEtran} 
\usepackage[utf8]{inputenc}
\usepackage{amsmath}
\usepackage{amssymb}
\usepackage{tabularx}
\usepackage{mathtools}
\usepackage{graphicx}
\usepackage{optidef}
\usepackage{amsfonts}
\usepackage{algorithm}
\usepackage{algorithmic}
\usepackage{caption}
\usepackage{subcaption}
\usepackage{amsthm}

\usepackage{xcolor}
\usepackage{breqn}
\usepackage{cuted}
\usepackage{multicol}

\ifCLASSINFOpdf
\else
\fi
\DeclareUnicodeCharacter{2212}{-}

\begin{document}
%
\title{Green Resource Allocation in Cloud-Native O-RAN Enabled Small Cell Networks}
%
%
%
\author{
     Rana M. Sohaib\IEEEauthorrefmark{2}, \IEEEauthorblockN{Syed Tariq Shah\IEEEauthorrefmark{1}, Oluwakayode Onireti\IEEEauthorrefmark{2}, Yusuf Sambo\IEEEauthorrefmark{2}, M. A. Imran\IEEEauthorrefmark{2}}
    \IEEEauthorblockA{
    \IEEEauthorblockA{\IEEEauthorrefmark{2}James Watt School of Engineering, University of Glasgow, Glasgow, UK\\ Emails: \{RanaMuhammad.Sohaib, Oluwakayode.Onireti, Yusuf.Sambo, Muhammad.Imran\}@glasgow.ac.uk}
    \IEEEauthorrefmark{1}School of Computer Science \& Electronic Engineering, University of Essex, Colchester, UK\\ Email: Syed.Shah@essex.ac.uk}
}



\maketitle

\begin{abstract}
In the rapidly evolving landscape of 5G and beyond, cloud-native Open Radio Access Networks (O-RAN) present a paradigm shift towards intelligent, flexible, and sustainable network operations. This study addresses the intricate challenge of energy efficient (EE) resource allocation that services both enhanced Mobile Broadband (eMBB) and ultra-reliable low-latency communications (URLLC) users. We propose a novel distributed learning framework leveraging on-policy and off-policy transfer learning strategies within a deep reinforcement learning (DRL)--based model to facilitate online resource allocation decisions under different channel conditions. The simulation results explain the efficacy of the proposed method, which rapidly adapts to dynamic network states, thereby achieving a green resource allocation. 

\end{abstract}


\IEEEpeerreviewmaketitle

\section{Introduction}
Open Radio Access Network (O-RAN) signifies a transformative approach to designing and implementing next-generation mobile networks. It seeks to break down the conventional barriers of proprietary network elements, advocating for a system where hardware and software components from various vendors work together seamlessly. Unlike traditional RAN setups where a single vendor might provide a closely knit hardware-software package, O-RAN advocates a modular and flexible framework \cite{I1, I2}. This framework divides radio network functions into standardised, interoperable units, empowering network operators with the ability to combine network elements from different suppliers. This ability not only encourages competition but also diminishes reliance on singular suppliers, with the added benefit of possibly reducing overall costs.
\begin{figure*}[t]
    \centering
        \includegraphics[width=\textwidth,height=0.6\textheight,keepaspectratio]{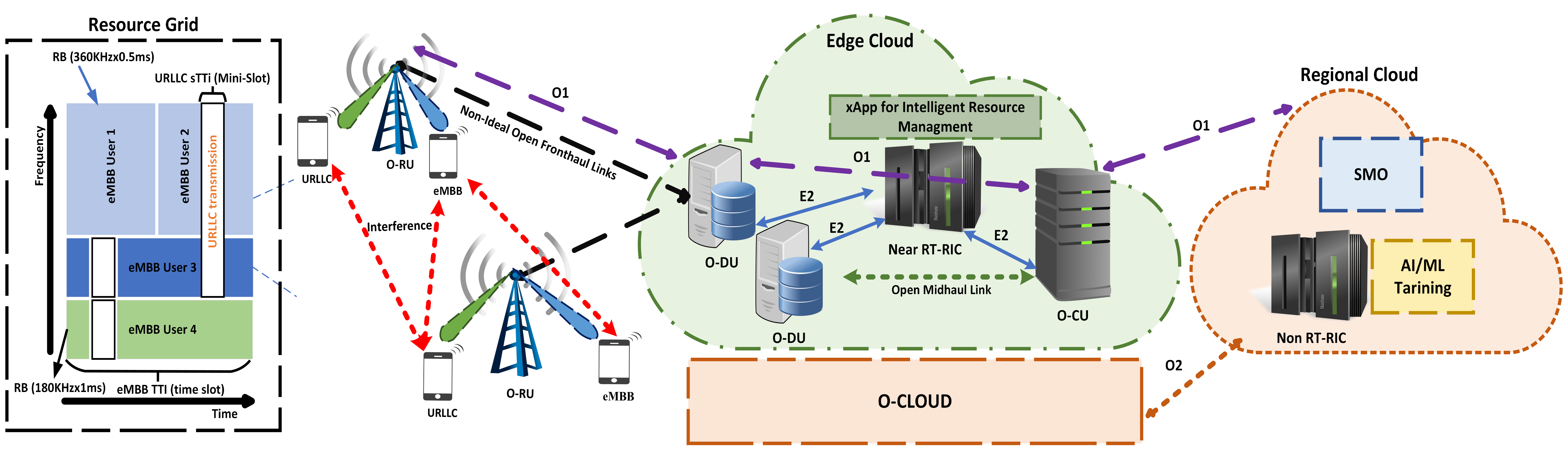}
    \caption{Considered Cloud Native O-RAN based Small Cell Network and Considered Resource Grid. }
    \label{fig1l}
\end{figure*}

Central to O-RAN's architecture are the Radio Unit (RU), the Distributed Unit (DU), the Centralized Unit (CU), and the RAN Intelligent Controller (RIC), each playing a pivotal role in ensuring the network's functionality and efficiency. The RU focuses on radio frequency processing, the DU on baseband processing, and the CU on managing higher-layer functions, all working in concert through open interfaces and standardised protocols \cite{I1}. This architecture ensures high flexibility, allowing network operators to tailor their infrastructure according to specific needs and preferences.
Moreover, introducing the RIC as a key component of O-RAN underlines the network's forward-thinking design. The RIC, with its near real-time and non-real-time variants, brings programmability and embedded AI capabilities into the network's core, enabling dynamic adaptation of network operations such as admission control, radio resource allocation, and power management \cite{I3, I5}. This adaptability is especially crucial in catering to the diverse and evolving demands of Beyond 5G (B5G) networks, which aim to significantly improve the network energy efficiency and serve a wide array of vertical industries with varying requirements for enhanced Mobile Broadband (eMBB) and Ultra-Reliable Low Latency Communications (URLLC) \cite{RW1}.

Since most of the energy usage in mobile networks is attributed to the radio access segment \cite{EE1}, efficient resource management schemes that ensure the network's energy efficiency (EE) and guarantee quality-of-service (QoS) for eMBB and URLLC users are essential \cite{I4, EE2}. More specifically, integrating EE considerations into the dynamic and complex resource management landscape of B5G O-RAN-based small cell networks is vital, particularly in the presence of URLLC and ENBB users. The study in \cite{RW1} proposes a method to enhance EE and ensure QoS in O-RAN by jointly optimising radio resource allocation and DU selection. The proposed method uses a linearised model solved by a Mixed Integer Linear Programming (MILP) solver, which outperforms traditional disjoint approaches by reducing the network energy consumption. In \cite{RW2}, the authors first introduce two machine learning-based closed-loop controls (CLCs) aimed at network traffic prediction and network slicing within the Non-RT and Near-RT RIC domains of O-RAN architecture. Following this, they propose an energy-efficient AI/ML pipeline tailored for deploying these CLCs. Their numerical results demonstrate the effectiveness of the proposed approach compared to fixed centralised and distributed deployments. In \cite{RW3}, the authors introduce an optimisation model for enhancing routing and energy efficiency in IAB networks. Using O-RAN's CLC framework, their model aims to reduce the number of active IAB nodes, ensuring minimal capacity for each user. They transform and solve a complex binary nonlinear problem into a linear one. Tested on Milan's network data, their method cuts RAN energy use by 47\% while ensuring user threshold capacity.
A distributed learning framework for resource allocation in multi-cell wireless systems catering to eMBB and URLLC users is presented in \cite{RW4}. Leveraging a Thompson sampling-based Deep Reinforcement Learning (DRL) algorithm within the O-RAN architecture, this approach enables real-time decision-making at the network edge by deploying execution agents at Near-RT RICs. Simulation results demonstrate that the algorithm is effective in meeting QoS requirements for diverse user groups and optimising resource utilisation in dynamic environments.

Unlike the above-mentioned works, this paper tackles the challenge of maximising network energy efficiency in a split-6 O-RAN-based small cell network, accommodating both eMBB and URLLC users. We formulate the energy efficiency maximisation problem in terms of achievable sum-rate and total consumed power. Our proposed approach, utilising on-policy and off-policy transfer learning strategies within a DRL-based clipped proximal policy optimization (PPO), seeks to optimise this objective function through physical resource blocks (PRB) allocation, radio resource puncturing decisions, and transmit power adjustments while ensuring the strict latency requirements of URLCC users. This method allows for a dynamic and intelligent allocation of network resources, aiming to enhance the overall energy efficiency of the network while meeting the diverse requirements of eMBB and URLLC services. Extensive simulation results show that the proposed approach achieves green resource allocation by rapidly converging to optimal resource distribution policies that maintain high EE, even under varying and unpredictable channel conditions. 


\vspace{-0.3cm}
\section{System Model}
We consider a downlink O-RAN scenario where we have $N$ numbers of DUs, defined as $\mathcal{N}=\left\{1, . . ., N \right\}$ are served by a single CU. There are $L$ numbers of RUs, defined as $\mathcal{L}=\left\{1, . . ., L \right\}$ served by each DU, where a RU $l\in L$ supports two different service users, a set of eMBB users $\mathcal{W}_{l,n}^{e}=\left\{1, . . ., W_{l,n}^{e} \right\}$, and URLLC users $\mathcal{W}_{l,n}^{u}=\left\{1, . . ., W_{l,n}^{u} \right\}$, as shown in Fig. 1. Each of these DUs is linked to the Near-RT-RIC, equipped with an intelligent scheduling xApp that determines the allocation of PRBs to all users, i.e., $\mathcal{\Bar{W}}=\{ 1, 2, \cdot\cdot\cdot, W\}$, where $W= \{W_{l,n}^{e} +W_{l,n}^{u}\}$ connected to their respective RUs. 
We positioned edge cloud servers at the near-RT-RIC, creating links to a regional cloud server at non-RT-RIC. As per \cite{b1}, the Non-RT RIC resides on the regional cloud server, while the Near-RT RICs are deployed on the edge cloud servers and every DU is connected to a solitary edge server. 
The radio spectrum resources in 5G New Radio (5G-NR) can be illustrated in both the frequency and time domains. These domains are divided into smaller sections called RBs, which comprise a total of $K$ radio resources. Every time slot is subdivided into $M$ mini-slots. Each RB is identified by a specific bandwidth represented as $B$. 
Typically, the eMBB service spans across multiple TTIs aiming to enhance spectral efficiency (SE). Nevertheless, the stringent latency demands necessitate immediate processing of incoming URLLC traffic.
To meet the rigorous requirements for low latency and high reliability in URLLC, the technique of puncturing eMBB slots is utilized. By allocating specific resources to URLLC traffic through puncturing eMBB slots, the prompt delivery of crucial information can be ensured with timeliness and reliability. The URLLC service is programmed to have a short TTI of 0.5 ms, whereas a longer duration of 1 ms is assigned for the eMBB service.
The prompt execution of URLLC transmission, involving the interruption of eMBB traffic, can significantly impact the capacity and reliability of the system. Consequently, this could lead to a reduction in the efficiency of the eMBB service.
Hence, a suitable structure is necessary to meet the QoS criteria.

\textbf{eMBB Rate:} Scheduling URLLC transmission promptly may compromise the throughput of the eMBB service. The puncturing decision variable \(\delta_{k,m}^{l,w}(t)\) is a binary variable that takes the value of 1 if the \(m^{th}\) mini-slot is punctured by the \(w^{th}\) URLLC user, applicable for all \(w\) in the set \(\mathcal{\Bar{W}}\) and \(l\) in the set \(\mathcal{L}\), at time \(t\). It is set to 0 in all other instances. 
The signal-to-noise-and-interference-ratio (SINR) of the eMBB user $w$ can be mentioned as follows
\begin{equation}
    \Omega_{l,k}^{e,w}(t)=\frac{p_{l,k}^{e,w}(t)g_{l,k}^{e,w}(t)}{\sum\limits_{\substack{l'\in\mathcal{L} \\ l'\neq l}}\underbrace{p_{l',k}^{e,w}(t)g_{l',k}^{e,w}(t)}_{\text{eMBB interference}}+\sum\limits_{\substack{l'\in\mathcal{L} \\ l'\neq l}}\underbrace{p_{l',k}^{u,w}(t)g_{l',k}^{u,w}(t)}_{\text{URLLC interference}}+\sigma^{2}},
\end{equation}
where $p_{l,k}^{e,w}(t)$, and $g_{l,k}^{e,w}(t)$ denotes the transmitted power and channel gain, respectively, of eMBB user $w$ of RU $l$ over RB $k$, and $\sigma^{2}$ denotes the noise power. The achievable rate for an eMBB user $w$ in RU $l$, utilizing RB $k$ during time slot $t$, can be determined as follows
\begin{align}
    r_{l,k}^{e,w}(t)=B\left(1-\frac{\sum_{m=1}^{M}\delta_{k,m}^{l,w}(t)}{M} \right)\log_{2}\big(1+\Omega_{l,k}^{e,w}(t) \big),
\end{align}
where the expression $\frac{\sum_{m=1}^{M}\delta_{k,m}^{l,w}(t)}{M}$ signifies the decline in eMBB rate due to puncturing. We presume that every RU allocates an RB for an individual user. 
Here we define an RB allocation decision variable \(\alpha_{w,k}^{l,n}(t)\) which is a binary variable that takes the value of 1 if RB \(k\) of Radio Unit \(l\) in DU \(n\) is assigned to the eMBB user \(w\), for all \(l\) in the set \(\mathcal{L}\) and \(n\) in the set \(\mathcal{N}\). It is set to 0 in all other cases, indicating that the RB is not allocated to that specific eMBB user. 
Therefore, the total throughput achieved by the eMBB user $w$ can be calculated as
\begin{align}
     r_{l,n}^{e,w}(t)=\sum\limits_{k\in\mathcal{K}}\alpha_{w,k}^{l,n}(t)r_{l,k}^{e,w}(t).
\end{align}

\textbf{URLLC Rate:}In order to minimise transmission latency, it's crucial to restrict the block length in URLLC. Shannon's capacity theorem becomes pertinent when handling an indefinite block length \cite{bs2}. The achievable throughput of URLLC with finite blocklength can be determined as
\begin{align}
     r_{l,k}^{u,w}(t)=&\sum\limits_{k\in\mathcal{K}}B_{k}\Big(\frac{\sum_{m=1}^{M}\delta_{k,m}^{l,w}(t)}{M} \Big)\Bigg[\log_{2}\big(1+\Omega_{l,k}^{u,w}(t) \big)\\ 
     &-\sqrt{\!\frac{\!D_{l,k}^{u,w}}{\!C_{l,k}^{u,w}(t)}}.Q^{\!-1}(\!x)\Bigg],\nonumber
\end{align}
where $C_{l,k}^{u,w}(t)$ refers to the symbols in each mini-slot, and $\Omega_{l,k}^{u,w}(t)$ represents the SINR for the URLLC user, as defined by
\begin{equation}
    \Omega_{l,k}^{u,w}(t)=\frac{p_{l,k}^{u,w}(t)g_{l,k}^{u,w}(t)}{\sum\limits_{\substack{l'\in\mathcal{L} \\ l'\neq l}}\underbrace{p_{l',k}^{u,w}(t)g_{l',k}^{u,w}(t)}_{\text{URLLC interference}}+\sum\limits_{\substack{l'\in\mathcal{L} \\ l'\neq l}}\underbrace{p_{l',k}^{e,w}(t)g_{l',k}^{e,w}(t)}_{\text{eMBB interference}}+\sigma^{2}}.
\end{equation}
Here, $D_{l,k}^{u,w}=1-\frac{1}{(1+\Omega_{l,k}^{u,w}(t))^{2}}$ represents the dispersion of the channel.
    

\section{Problem Formulation}
The coexistence of URLLC users alongside eMBB traffic adds extra pressure on eMBB transmissions, potentially leading to a violation of the minimum QoS requirements.
It is presumed that the URLLC traffic produces shorter packet fragments. The arrival rate of packets in mini-slot $m$, where $m$ is within the set $\mathcal{M} = \{1, ..., m, ..., M\}$, during TTI $t$, conforms to a Poisson point process (PPP) distribution and is denoted as $\beta_{m}(t)$.
Furthermore,  the aggregate count of URLLC packets received within the TTI $t$ can be computed as 
\begin{align} \label{phi}
    \beta(t)=\sum_{m\in \mathcal{M}}\beta_{m}(t).
\end{align}
In light of considering non-ideal FH, it is necessary to incorporate the round-trip delay of HARQ retransmission into the URLLC service's outage probability. This adaptation becomes essential to appropriately account for the extra delay introduced by HARQ retransmissions, thereby incorporating both the initial transmission and potential retransmissions into the determination of the outage probability. Consequently, the reliability equation, which encompasses non-ideal FH and HARQ, can be formulated as follows 
\begin{align}\label{eq:out}
   \Pr\left[r_{l,k}^{u,w}(t+\lambda_{RTT})\leq \varrho\beta(t)\right]\leq \sigma_{u},\hspace{0.75cm}\forall l\in L,
\end{align}
where $\lambda_{RTT}$ and $\varrho$ refer to the round-trip delay of HARQ retransmission in TTI units and URLLC packet size, respectively. This adjustment reflects the shift towards evaluating the system's reliability at time $(t+\lambda_{RTT})$ rather than solely at time $t$, thereby accounting for the timing of reliability assessment.
To characterise the success probability for each HARQ transmission, BER-based methodologies are utilised \cite{bhar1, bhar2}. At time $(t+\lambda_{RTT})$, the probability of successful transmission is estimated by considering the cumulative distribution function (CDF) of the bit error rate (BER). Following this, a Bernoulli experiment is conducted to ascertain the success or failure of the transport block (TB) transmitted to each user based on these estimated success probabilities. Such a modelling approach facilitates the assessment of delays' influence on HARQ transmission outcomes and, consequently, on the overall data rate for URLLC services.
We characterise the energy efficiency (EE) of the system by calculating the ratio of the aggregate data rate to the overall power consumption, specifically
\begin{align}
   \zeta_{EE}^{n}(t)= \frac{\sum_{l\in\mathcal{L}}\sum_{w\in \mathcal{\Bar{W}}}\{r_{l,n}^{e,w}(t) + r_{l,k}^{u,w}(t) \}}{\sum_{l\in\mathcal{L}}\sum_{w\in \mathcal{\Bar{W}}}\sum_{k\in\mathcal{K}}p_{l,k}^{w}(t) + L\cdot P_{RU}(t) + P_{DU}^{n}},
\end{align}
where $L\cdot P_{RU}(t)$, and $P_{DU}^{n}$ refers to the amount of total circuit power consumed by $L$ RUs and $n^{th}$ DU at each TTI, respectively.
Consequently, we formulate an objective function aimed at enhancing the EE of the network as follows
\begin{subequations}\label{eq:op}
\small 
\begin{align}
\textbf{P}:\max_{\alpha, P, \delta}&\left\{\zeta_{EE}^{n}(t)\right\} \label{eq:obf} \\
\text{s.t.}&\sum_{w\in \mathcal{W}_{l,n}^{e}}\!\alpha_{w,k}^{l,n}(t)\leq1, \forall k\in\mathcal{K}, l\in\mathcal{L}, n\in\mathcal{N} \label{eq:c1}\\
&\sum_{w\in \mathcal{W}_{l,n}^{u}}\!\delta_{k,m}^{l,w}(t)\leq1, \forall k\in\mathcal{K}, l\in\mathcal{L}, n\in\mathcal{N}\label{eq:c2}\\
&\sum_{m\in \mathcal{M}}\!\delta_{k,m}^{l,w}(t)\leq M, \forall k\in\mathcal{K}, l\in\mathcal{L}\label{eq:c3}\\
&\Pr\left[r_{l,k}^{u,w}(t+\lambda_{RTT})\leq \varrho\beta(t)\right]\leq \sigma_{u}, \forall l\in L\label{eq:c4}\\
&\sum_{w\in\mathcal{W}_{l,n}^{e}}r_{l,n}^{e,w}(t)\leq \Bar{r}_{e}\label{eq:relc} \\
&\sum_{w\in\mathcal{W}_{l,n}^{e}}\sum_{k\in\mathcal{K}}p_{l,k}^{e,w}(t)\leq P_{max}, \forall l\in\mathcal{L}\label{eq:c5}\\
&p_{l,k}^{e,w}(t)\geq 0, \forall w\in\mathcal{W}^{e}, k\in\mathcal{K} \label{eq:c6}\\
&\alpha_{w,k}^{l,n}(t)\in\{0,1\}, \forall w\in\mathcal{W}^{e}, k\in\mathcal{K}\label{eq:c7}\\
&\delta_{k,m}^{l,w}(t)\in\{0,1\}, \forall w\in\mathcal{W}^{u}, k\in\mathcal{K}\label{eq:c8}
\end{align}
\end{subequations}
where (\ref{eq:obf}) denotes the objective where we look to maximise the EE of the system. Constraint (\ref{eq:c1}) sets the limit on the allocation of resources for eMBB, ensuring that each RB is assigned to only one user.
Whereas (\ref{eq:c2}) ensures that at any given time slot, only one URLLC user is permitted to puncture a specific mini-slot within an RB. Constraint (\ref{eq:c3}) states that the number of punctured mini-slots must not exceed the total count of mini-slots available. Constraint (\ref{eq:c4}) establishes the reliability requirement for URLLC. Constraint (\ref{eq:relc}) refers to the eMBB reliability. Constraints (\ref{eq:c5}) and (\ref{eq:c6}) outline the power allocation restrictions for eMBB. Similarly, the constraints (\ref{eq:c7}) and (\ref{eq:c8}) delineate the constraints concerning resource allocation visibility limitations.
\section{Proposed Distributed Learning Approach}
The optimisation problem in (\ref{eq:op}) is categorised as NP-hard. To address this, we conceptualise the problem as a Markov decision process (MDP) involving $N$ agents.

\subsubsection*{state space}
We make the assumption that each edge cloud server, functioning as an agent, receives only its own distinct state, focusing particularly on the data pertaining to users within the corresponding cell. The set of states can be described as $S_{n}=\left\{S_{1,n}, S_{2,n}, . . . , S_{l,n}, . . ., S_{L,n} \right\}$. 
The state space includes the channel information of both eMBB and URLLC users, as well as the traffic details at time slot $t$, and it can be represented as $s_{l,n}(t)=\left\{g_{l,n}^{e}(t), g_{l,n}^{u}(t), \varrho_{l,n}(t), W_{l,n}^{e}, W_{l,n}^{u}\right\}$.  
\subsubsection*{Action}
The set of action space can be described as $A=\left\{A_1, A_2, . . . , A_{n}, . . ., A_{N} \right\}$. Each agent makes decisions regarding the selection of eMBB RB denoted by $\alpha$, the allocation of eMBB power represented by $P$, and the scheduling of URLLC denoted by $\delta$. 
\subsubsection*{Reward}
We formulate the global reward function, taking into account the specifications of URLLC users. The reward function can be expressed as follows
\begin{align}\label{eq:rewa}
r(t)=&\overbrace{{\zeta_{EE}^{n}(t)}}^{I} -\Psi(t)\bigg(\overbrace{ r_{l,k}^{u,w}(t+\lambda_{RTT}) - \varrho\beta(t) }^{II}\bigg) - r_{l,n}^{e,w}(t)
\end{align}
where we incorporate the time-dependent weighting coefficient, denoted as $\Psi(t)$, to guarantee the reliability constraint of URLLC. It can be updated as follows
\begin{align}
    \Psi(t+1)=\max\left\{\Psi(t)+\Phi(t)-\sigma_{u},0\right\},
\end{align}
where $\Phi(t)$ represents the transmission error probability. In the first part, we aim to maximise the EE of the network, while the second part represents the constraint for URLLC. 
\subsection{Multi-agent DRL Framework with Transfer Learning}
In this setup, DUs acting as edge servers are interconnected with near-RT-RIC. The intelligent scheduling xApp, operating within the near-RT-RIC, is responsible for the dynamic allocation of PRBs to users. Each edge server serves as an autonomous agent in this decentralised framework, capable of making localised decisions to optimise resource allocation and network performance. Meanwhile, the core training module resides within the regional cloud server. 
Offline training of a fully connected neural network (NN) model is conducted at the non-RT-RIC in the regional cloud, utilising data gathered from all edge agents. This trained global model is then communicated to the DRL agents situated at the near-RT-RIC deployed at the edge servers. The objective of training the global model is to maximise a designed global reward function in (\ref{eq:rewa}), thus optimising the network's overall performance and efficiency. In DRL, NN is commonly employed to parameterise the function of the critic and actor.
In this work, we adopt the policy gradient-based approach to update the actor part. We utilise the clipped PPO, where the objective of the policy function is set to be the minimum value between the standard surrogate objective and a clipped objective. It can be formulated as follows
\begin{align}\label{eq:onp}
    \nabla_{clip}(\theta)=\mathbf{E}\bigg[\min \Big(\mu_{\theta}(s_{t}, a_{t})\cdot\gamma, \text{clip}_{\epsilon}(\mu_{\theta}(s_{t}, a_{t}) ) \Big)\cdot\gamma \bigg],
\end{align}
where the $\theta$ defines the policy $\pi$, while $\gamma$ represents the estimates of advantages.  Additionally, $\mu_{\theta}(.,.)$ denotes the likelihood ratio and can be expressed as 
\begin{align}\label{eq:onpp}
    \mu_{\theta}(s_{t}, a_{t})=\frac{\pi_{\theta}(a_{t}| s_{t})}{\pi_{\Bar{\theta}}(a_{t}|s_{t})}.
\end{align}
Furthermore, the clip function limits the range of $\mu_{\theta}(.,.)$ to a specific interval. We present the implementation of policy updates by employing Clipped PPO and a solitary master policy for knowledge transfer. In actor-critic RL, the update process for the critic involves employing conventional supervised regression. This method mirrors the approach utilised in Clipped PPO. However, two fundamental concepts are the foundation for our actor update: on-policy transfer learning for allocating RBs at near-RT-RIC and off-policy transfer learning in non-RT-RIC. This approach allows for more immediate feedback and adaptation to changing network conditions. On-policy methods directly update the policy based on experiences collected during interaction with the environment, 
The transfer of on-policy utilises a technique known as policy distillation \cite{btr1}.
In the off-policy transfer process, we fine-tune the actor by defining an objective $\nabla_{off}$. This is achieved by employing an advantage-based experience selection technique. The suggested method for experience selection aims to choose samples that exhibit a strong semantic connection to the target task rather than solely focusing on similarity.
\subsection{On-policy Transfer Learning at near-RT-RIC}
To enhance the initial performance of DRL agents, we employ a policy distillation method for on-policy transfer learning \cite{btr1}. It involves a two-step process: first, training a master policy that performs well on a source task, and then distilling this policy into a learner policy that can operate in the target task environment.
On-policy transfer learning at near-RT-RIC involves adapting and optimizing the policy based directly on the experiences gained during the interaction with the environment in a near-real-time fashion. This method continuously updates the policy based on the latest actions and observed rewards, aiming to improve the decision-making process for resource allocation.
The core concept revolves around utilising an additional loss function that incentivises the learner policy to closely resemble the master policy along the trajectories sampled by the learner. The learner learns by imitating the master policy. This can be done by minimizing the difference between the state-action value functions of the master and the learner.
Considering a master policy denoted as $\pi_{master}$, we introduce an additional loss function as follows
\begin{align}
    \nabla_{add}(\theta) = Z\bigg(\pi_{master}(a|s) || \pi_{\theta} (a|s)\bigg),
\end{align}
where $Z(.||.)$ refers to the cross-entrpoy.
The policy distillation process incorporates the loss mentioned in (\ref{eq:onp}) into the clipped PPO objective function as follows
\begin{align}
    \nabla_{on}^{j}=\nabla_{clip}(\theta) - \eta_{j}\nabla_{add}(\theta),
\end{align}
where $\eta_{j}$ denotes the weighting factor. It helps to improve the initial performance of the DRL agents. 
\subsection{Off-policy Transfer Learning at non-RT-RIC}
Off-policy transfer learning in a non-RT-RIC can be initiated when a sufficient amount of data has been collected from the edge servers and stored in the replay buffer. The first step is to collect training data from various edge servers or agents following the master policy $\pi_{master}$. This data typically includes observations of the environment, actions taken by the agents, and corresponding rewards from the target task. The collected data should be stored in a replay buffer.
Considering that the samples originate from a distinct distribution, we refrain from utilising them to update the state value (critic) network directly. However, to enhance sample efficiency, during the policy (actor) network update, we adopt a selective approach. We choose transitions based on their associated advantage values and exclusively incorporate samples exhibiting advantages surpassing a predetermined threshold. Additionally, considering that the master policy was employed during data collection, we adjusted the objective function for off-policy learning. This entails omitting the additional cross-entropy loss and substituting $\pi\Bar{\theta}$ with $\pi_{master}$ in (\ref{eq:onpp}), ensuring a more effective transfer of knowledge from the teacher policy to the learning process. It can be expressed as follows
\begin{align}\label{eq:onp2}
    \nabla_{off}(\theta)=\mathbf{E}\bigg[\min \Big(\nu_{\theta}(s_{t}, a_{t})\cdot\Bar{\gamma}, \text{clip}_{\epsilon}(\nu_{\theta}(s_{t}, a_{t}) ) \Big)\cdot\Bar{\gamma} \bigg].
\end{align}
Hence, as advantages are computed based on the reward function specific to the target task, state-action transitions exhibiting high values are considered favourable for transfer learning. This approach helps to enhance the sample efficiency. 
\section{Performance Evaluation}
We consider four DUs and eight RUs in our network model, where each DU serves two RUs, and each RU offers coverage to an area spanning 250 square meters and manages both eMBB and URLLC users. The eMBB users continuously produce data with full-buffer traffic, whereas URLLC users generate traffic characterised by a Poisson distribution, with an arrival rate represented by $\beta$.
The simulation parameters are defined in Table \ref{tab:1}.

\begin{table}[t]
\caption{Simulation Parameters} 
\centering
\label{tab:1}
\begin{tabularx}{\columnwidth}{|X|X|}
\hline
\textbf{Parameter} & \textbf{Value} \\ 
\hline\hline
Frame duration & 10 ms \\ 
\hline
No. of mini-slots in each TTI & 7 \\ 
\hline
sub-carrier spacing & 15 Khz \\ 
\hline
No. of OFDM symbols/TTI & 14 \\ 
\hline
OFDM symbols/mini-slot & 2 \\ 
\hline
Bandwidth & 20 MHz \\ 
\hline
URLLC packets length & 32 Bytes  \\ 
\hline
RB Bandwidth & 180 kHz \\ 
\hline
RU Transmit power & 38 dBm \\ 
\hline
DU Power & 200 W \\
\hline
Pathloss Model & $120.8 + 37.5 \log_{10}(d)$ \\ 
\hline
Actor learning rate & $10^{-5}$ \\ 
\hline
Critic learning rate & $10^{-3}$ \\ 
\hline
\end{tabularx}
\end{table}
\begin{figure}[t]
     \centering
\includegraphics[width=0.5\textwidth]{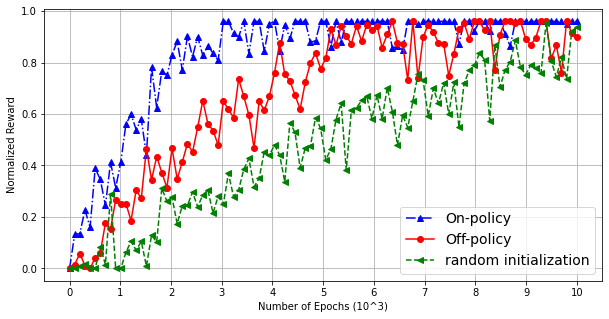}
\caption{Learning comparison}
\label{fig:fig2}
\end{figure}
Fig. \ref{fig:fig2} illustrates the convergence of the on-policy and off-policy transfer learning models in contrast to a baseline model with random initialisation. Here, the source and target tasks have different wireless conditions, implying that channel characteristics such as path loss, shadowing, and fading patterns vary significantly between the tasks. The on-policy transfer learning approach demonstrated a robust convergence pattern, as shown in the figure. The sustained high reward of the on-policy method suggests that the model quickly identifies and adheres to resource allocation strategies that are both effective and EE. 
Conversely, the off-policy method portrayed a more volatile convergence trajectory. The results indicate that the on-policy transfer learning method implemented within the near-RT-RIC is more suitable for environments requiring quick and reliable convergence, such as achieving the EE while balancing the dual demands of eMBB reliability and URLLC constraints. The off-policy approach, while valuable for its capacity to leverage diverse experiences, might necessitate further optimisation to achieve comparable levels of efficiency within the same operational constraints. The random initialisation serves as a baseline to contextualise the learning efficiency of the transfer learning methods. The convergence plot presents compelling evidence of achieving green resource allocation where our proposed learning model ensures that EE resource optimization are not compromised when the network transitions from one set of channel conditions to another.

\begin{figure}[t]
     \centering
\includegraphics[width=0.5\textwidth]{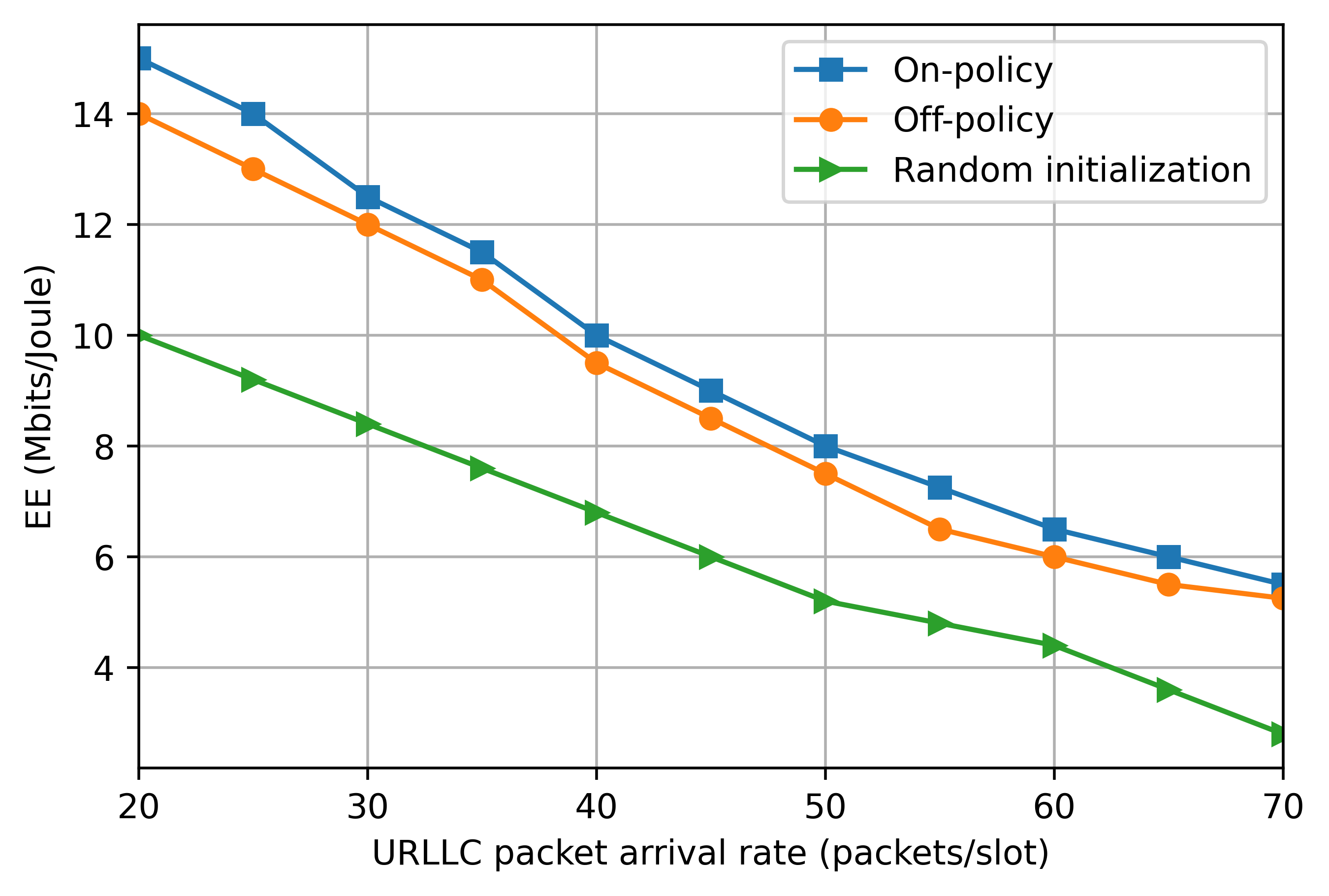}
\caption{URLLC impact on EE of the system}
\label{fig:fig5}
\end{figure}
In Fig. (\ref{fig:fig5}), we present the EE performance of on-policy and off-policy transfer learning methods, in addition to a baseline random initialisation approach, under varying URLLC packet arrival rates. The EE of all methods naturally declines as the URLLC packet arrival rate increases, indicating higher energy demands to handle the additional packet traffic. The on-policy method exhibited superior energy efficiency across all URLLC packet arrival rates. The on-policy method, in particular, demonstrates robustness in maintaining higher energy efficiency, reflecting its potential for sustainable network operations, especially in scenarios characterised by high URLLC traffic under varying channel conditions. The off-policy method, while less efficient than the on-policy, still significantly outperforms the random initialisation baseline, emphasising the value of structured learning algorithms in energy-conscious network management.
\section{Conclusion}
In this work, we have addressed the resource allocation challenges with a keen focus on multiplexing of eMBB and URLLC services. 
Through the formulation of an optimisation problem and the development of a distributed learning framework, particularly leveraging a transfer learning approach, we have presented a solution capable of making online resource allocation decisions under different channel conditions. Our incorporation of transfer learning approaches, both on-policy and off-policy, within the framework of O-RAN network architectures has demonstrated a marked improvement in energy efficiency metrics. The simulation result shows that the on-policy transfer learning method, in particular, showcased its prowess by not only converging rapidly but also by optimising the EE across varying levels of URLLC packet arrival rates. Our work contributes critical insights into the optimisation of resource utilisation, embracing the dual imperatives of dynamic wireless environment adaptability and sustainable network management. Our findings advocate for the integration of intelligent transfer learning mechanisms that can dynamically adapt to network conditions while significantly reducing the carbon footprint of wireless communications.

\ifCLASSOPTIONcaptionsoff
  \newpage
\fi


\bibliographystyle{IEEEtran}
\bibliography{ref}


%








\end{document}